# Building Effective Safety Guardrails in AI Education Tools

Hannah-Beth Clark, Laura Benton, Emma Searle, Margaux Dowland, Matthew Gregory, Will Gayne and John Roberts.

Oak National Academy, UK
{hannah-beth.clark,laura.benton,emma.searle,eng.mdowland,matthew.gregory,will.gayne,john.roberts}@thenational.academy

**Abstract.** There has been rapid development in generative AI tools across the education sector, which in turn is leading to increased adoption by teachers. However, this raises concerns regarding the safety and age-appropriateness of the AI-generated content that is being created for use in classrooms. This paper explores Oak National Academy's approach to addressing these concerns within the development of the UK Government's first publicly available generative AI tool - our AI-powered lesson planning assistant (Aila). Aila is intended to support teachers planning national curriculum-aligned lessons that are appropriate for pupils aged 5-16 years. To mitigate safety risks associated with AI-generated content we have implemented four key safety guardrails: (1) prompt engineering to ensure AI outputs are generated within pedagogically sound and curriculum-aligned parameters; (2) input threat detection to mitigate attacks; (3) an Independent Asynchronous Content Moderation Agent (IACMA) to assess outputs against predefined safety categories; and (4) taking a human-in-the-loop approach, to encourage teachers to review generated content before it is used in the classroom. Through our on-going evaluation of these safety guardrails we have identified several challenges and opportunities to take into account when implementing and testing safety guardrails. This paper highlights ways to build more effective safety guardrails in generative AI education tools including the on-going iteration and refinement of guardrails, as well as enabling cross-sector collaboration through sharing both open-source code/datasets and learnings.

**Keywords:** AI, education, safety guardrails, moderation.

## 1 Description of the AIED Implementation

## 1.1 Background

Over the last five years, the emergence of accessible generative AI has led to an explosion of tools with integrated AI within the education sector. Decisions on which content is appropriate and should or shouldn't be taught in schools is currently set out by the UK Government through the national curriculum objectives and guidance (Department for Education, 2019). School leaders interpret and implement this within their schools by creating their own school-based curricula and policies, informing teachers' decisions and lesson planning.

There has been a massive increase in AI usage amongst teachers, since the launch of accessible generative AI tools like ChatGPT in November 2022 with almost 70% of teachers now using it in their work and one in ten secondary teachers using it during lessons (Wespieser, 2024). The content produced through these tools should align with the Government and schools' expectations of appropriate content. This is challenging due to the susceptibility of AI tools to bias, hallucinations, and inaccuracies and its ability to produce potentially harmful content (Farquhar et al., 2024; MIT Management, n.d.). Generative AI's tendency to produce different responses each time it is asked a question makes testing the outputs of generative AI tools more challenging.

This area of technology is developing incredibly rapidly, and both research and legislation must keep up due to the associated risks of using the content produced in schools with children. The UK Government recently published guidance on the safety expectations for using Generative AI (Department for Education, 2025) in products used in education settings to support this work. However, this is guidance rather than legislation. It is up to individual organisations to interpret and implement effective guardrails within their product based on this guidance. They may also choose to evaluate the effectiveness of their guardrails, but this is not currently compulsory.

## 1.2 Oak National Academy and Aila

Oak National Academy[1] was set up as an Arm's Length Body (ALB) of the Department for Education in England after the COVID-19 pandemic to improve pupil outcomes and close the disadvantage gap by providing teachers with high-quality curricula and resources. Over the last two years, we have been working with curriculum partners to create over 10,000 lesson resources for UK teachers of pupils aged 5-16 years, designed by subject experts and aligned to the national curriculum for England and exam boards. To ensure consistency across multiple creators, we have distilled leading pedagogy and cognitive science research into a detailed curriculum and lesson design rubric. Specific principles such as '*the prior knowledge starter quiz activates the necessary knowledge required to access the lesson*' or '*explanations are framed around small steps*' are incorporated into the rubric to guide lesson creation. All expert teacher creators use this same rubric to review and ensure lessons are consistently high-quality.

Using guidance from a number of sources, including the UK Government, the UK Council for Child Internet Safety, the British Board of Film Classification (BBFC), the BBC and the Pan European Game Information (PEGI) (BBC, n.d.; BBFC, 2024; Department for Education, 2024; Department for Science, Innovation and Technology, 2024; PEGI, n.d.; UKCCIS, 2015), Oak has also created a content guidance rubric that lesson creators can use to highlight to teachers where lessons may require sensitivity (such as lessons on drugs, cyberbullying, domestic abuse, sex education and the Holocaust).

---

[1] https://www.thenational.academy/

This high-quality corpus of content and codification of our pedagogical quality definition put us in a unique position to innovate with AI, and we launched an AI lesson planning assistant (Aila) in September 2024. We have designed Aila to co-plan lessons with teachers, supporting them in creating high-quality lesson plans and resources for their classrooms. These are aligned with the national curriculum and can be personalised to their pupils based on locality, needs or interests. More details of our work to ensure the quality of our AI-generated content can be found in our recently published paper with MIT Open Learning (Clark et al., 2025).

Aila is the first government-funded, publicly available generative AI tool in the UK. It has been designed to produce content for classrooms with pupils aged 5-16 years, so safety is imperative. Whilst GPT-4o (the underlying large language model Aila currently uses) has in-built moderation and filtering mechanisms (OpenAI, n.d.), given the education context of use, we did not feel this provided a sufficient safety net. Therefore, in addition to our existing content guidance for expert teacher content creators (informed by the national curriculum), we identified several other key AI safety guidelines which were available at the time of Aila's design (Mazeika et al., 2024; Ofsted, 2024; UNESCO, 2022). Based on these guidelines, we have incorporated four additional safety guardrails, including explicit instructions within Aila's prompt, input threat detection, an independent asynchronous content moderation agent, and the teacher as the human-in-the-loop. We discuss these in further detail below.

## 2 Reflection of the Challenges and Opportunities associated with the Implementation

### 2.1 Our Four Stages of Safety

**Prompt Engineering.** We have written an extensive prompt (which is part of our open-source codebase), codifying Oak's definition of best practices in pedagogy and cognitive science according to the latest research in these fields. Within this, we have also outlined what is and isn't appropriate for Aila to create. This includes considering the pupil's age, subject, and topic being planned and working to ensure that content is aligned with the national curriculum and appropriate for use in schools. Our prompt will steer Aila in producing lessons suitable for use in classrooms. For example, within a lesson on cannabis, Aila will suggest age-appropriate content based on the Relationships, Sex and Health Education (RSHE) curriculum (Department for Education, 2021) for Years 7-11 (ages 11-16 years), covering the effects of using cannabis, as well as the legal and societal impact of use.

**Input Threat Detection.** We have implemented an input threat detection layer, which monitors user inputs to check for inappropriate, manipulative, or malicious inputs, predominantly through prompt attacks, in order to reduce the risk of jailbreaking. Any input flagged as a threat will not be sent to the large language model (LLM) but will be flagged to our internal moderation systems, and any users who are repeatedly detected will be blocked.

**Independent Asynchronous Content Moderation Agent (IACMA).** We have built an independent, context-unaware AI agent to assess Aila's outputs. We used Oak's established framework for assessing and tagging content that may require additional guidance when designing our 'content guidance' categories, alongside the HarmBench evaluation framework (Mazeika et al., 2024), to define our 'toxic' categories.

Our IACMA has no context—either of what Aila has been instructed to do or what the user inputted—so this cannot influence the moderation agent's judgements. The agent categorises content from Aila into three categories, each composed of further sub-categories.

- **Safe**: content is appropriate for use in classrooms with the specified age group;
- **Content guidance**: content is appropriate but may require additional considerations or sensitivity from the teacher when delivering. The topics include physical or practical activities, upsetting/sensitive content or language, discussion of discriminatory behaviour or language, nudity or sexual content, violence or crime.
- **Toxic**: inappropriate for use in classrooms (highly sensitive or harmful). This includes content on encouraging harmful behaviour (inc. self-harm) or illegal activity, creation of weapons or harmful substances or encouragement of violence.

Our IACMA scores each section of content produced by Aila with a justification against the predefined subcategory descriptions. We use a 5-point Likert scale to evaluate the severity of content within each category.

A score of 5 indicates 'no concern'. Within the 'content guidance' category a score lower than this triggers the appropriate guidance being shown to the teacher. For any score lower than 5 in a 'toxic' category, the lesson will be blocked, and the user session will be ended. In this instance, an alert is sent to our AI team to notify them of toxic use, it is recorded for reporting purposes, and the lesson is no longer accessible to the user. Repeated attempts to plan toxic lessons will result in a user being blocked from using Aila to plan further lessons.

As we use AI to evaluate and make decisions on AI-generated content, the decision is not always consistent. Aila is being used to produce content for pupils as young as five, so the safety of content is imperative. We have designed our IACMA to be oversensitive. This may result in the 'content guidance' warning being shown and lessons being blocked as 'toxic' when not necessary, but we feel that this conservative approach is appropriate. As we continue to evaluate and iterate, we will improve the specificity of this agent to ensure that we minimise false positives that may affect user experience while maintaining high sensitivity, and therefore safety.

**Human-in-the-loop.** Our tools and resources are designed for teachers who know their curriculum and their pupils; therefore, our final safety guardrail is the human teachers themselves. We designed Aila so that the teacher leads the lesson planning process, with the AI suggesting content that they may want to include in their lesson. It is intentionally designed to always have a teacher between the AI-produced content and the pupils. Therefore, they are expected to review the content before and after download to ensure it is aligned with their school's policies and appropriate for their pupils.

## 2.2 Evaluating the effectiveness of our safety guardrails

**Pre-launch evaluation.** Before launching Aila, we undertook a rigorous testing process to evaluate the initial effectiveness of our safety guardrails. Using the corpus of Oak lessons created by our expert curriculum partners, which had already been assessed for sensitive content against these categories, we could test and align our IACMA, ensuring that the judgments it made matched those made by the humans creating the lesson. We also participated in several red-teaming exercises to ensure our tool aligned with the HarmBench standardised evaluation framework.

**Post-launch evaluation.** In addition, we have created an illustrative lesson dataset using Aila to plan lessons in bulk on a number of representative topics that came out of our initial evaluations. We have been able to use these, in addition to the 45,000 lessons that users have created post-launch, to further test our safety guardrails.

We have systematically reviewed the safe, content guidance and toxic lessons from both the user-created and our illustrative lessons with technical and subject experts. This process has highlighted a number of edge cases, such as lessons based on current affairs that occurred after the latest data that the underlying LLM has been trained on (e.g. the 2024 American election and recent conflicts) and lessons on topics outside of the national curriculum but still taught in schools (e.g. lessons on specific special educational needs or outdoor skills).

## 2.3 Challenges and Opportunities

Through our testing, we have identified several challenges and considerations to take into account when testing safety guardrails.

**Use of an illustrative data set.** We created an illustrative dataset using Aila to create complete lessons without user input. This has allowed us to test our safety guardrails on lessons at scale and in different contexts, for example, testing the behaviour of Aila and our moderation agent for a topic when planned for different aged pupils. When using a large data set created in this way, the impact of teacher input is not included, which meant we could not solely rely on this data for testing our safety guardrails.

Due to Aila's design, lessons created without user input will be the most school and age-appropriate version of a lesson. This means that many user-created lessons on the same topic may have received a different moderation category to those in our illustrative dataset as input from the user could steer the content produced away from being national curriculum aligned and appropriate for the classroom. This highlights the importance of ensuring that real usage data is (where possible) included in the testing process.

**Evaluation process.** Aila's moderation agent makes multiple checks during the lesson, checking each lesson section as it is produced and modified. When running the IACMA on the illustrative and user-created lesson sets (where the whole lesson was already created), the moderation category was often misaligned with the original moderation. This was because, in a real usage scenario, the initial sections of a lesson may have received a 'content guidance' or 'toxic' flag before the rest of the content was produced. As more content was created, and if this content was appropriate for use in a classroom, the classification given to the lesson would change. For example, a lesson titled 'weapons of mass destruction' may initially be flagged as being 'toxic' (and be blocked) at the point of initiation. When Aila is given a completed lesson to moderate, the content within the lesson may be for a Year 11 religious studies lesson that focuses on the impact, ethical arguments and religious views on weapons of mass destruction - appropriate for the GCSE religious studies specification (AQA, 2016). The moderation given to the completed lesson may be downgraded to 'content guidance'. In real usage, this lesson may have received a 'toxic' moderation due to the title and never reached a point where the content was produced, meaning the category given to the real-life lesson and the lesson in the testing dataset would have received different moderation categories.

To overcome these challenges, we have tested both the illustrative and user-created lesson sets (to overcome the first challenge outlined) and run the moderation in chunks that replicate the process used by the IACMA. This has provided a more representative data set.

## 3     Description of future steps

### 3.1     Future steps

This paper has focused on the initial steps we have taken to ensure that Aila produces safe and appropriate content for use in schools. Further work will focus on our steps to address the edge cases (examples mentioned above) and additional tests on these changes.

Generative AI is evolving rapidly, so our approach to safety must continue to be responsive and evolve based on emerging needs. Aila is currently unimodal (with only text as the input and output). As it develops to incorporate other mediums (including, for example, imagery and audio), our safety processes will need to evolve to include these.

### 3.2     Implications for other organisations

The use of generative AI in primary and secondary education (aged 5-16 years) is still in its infancy. As such, limited models of effective safety guardrails and datasets exist to test these. To date, this work has resulted in two key contributions to the wider community:

1. Our code, including our prompt, is open source[2], and our content is available under an open licence (OGL). Other organisations can freely access the work we have done to build Aila and specifically our moderation agent. We do this with a view to enable other organisations in the sector to have a headstart with building effective safety guardrails for their products.
2. We have created an illustrative dataset of over 1000 lessons (used as part of our guardrail evaluation) representative of the types of lessons created by users. We hope that by making this available to other organisations in the sector, we will enable them to test their AI guardrails' effectiveness using a context-specific dataset.

AI has enormous potential within the education sector to support students and teachers. The emergence of new, innovative projects and tools within this space presents a huge opportunity, but safety must be considered at the forefront of any developments. It is imperative that organisations developing generative AI tools think carefully about the audience, use, and risks of their tool in educational contexts, build appropriate safety guardrails, and evaluate these effectively, and for governments to support organisations in doing this.

This paper presents our initial work in this area, but this is just the beginning. We will continue to reflect and iterate on these processes as we develop Aila further.

**Acknowledgements.** We are grateful to all of the Oak National Academy curriculum partners and the wider team who have enabled the creation of our corpus of lessons, enabling our work in AI. We would also like to thank the AI squad for contributing to our work on quality and safety within Aila and to teams from the Department for Education and the wider UK Government for their support of the project.

**Disclosure of Interests.** The authors have no competing interests to declare relevant to this article's content.

---

[2] https://github.com/oaknational/oak-ai-lesson-assistant